# Thickness-dependent spin bistable transitions in single-crystalline molecular 2D material

John Koptur-Palenchar[1], Miguel Gakiya-Teruya[2], Duy Le[3], Jun Jiang[1], Rui Zhang[1], Xuanyuan Jiang[1], Hai-Ping Cheng[1], Talat S. Rahman[3], Michael Shatruk[2*], Xiao-Xiao Zhang[1*]

[1]Department of Physics, University of Florida, Gainesville, FL, USA

[2]Department of Chemistry, Florida State University, Tallahassee, FL, USA

[3]Department of Physics, University of Central Florida, Orlando, FL, USA

**Abstract:**

The advent of two-dimensional (2D) crystals has led to numerous scientific breakthroughs. Conventional 2D systems have in-plane covalent bonds and a weak out-of-plane van-der-Waals bond. Here we report a new type of 2D material composed of discrete magnetic molecules, where anisotropic van-der-Waals interactions bond the molecules into a 2D packing. Through mechanical exfoliation, we can obtain single-crystalline molecular monolayers, which can be readily integrated into other 2D systems. Optical spectroscopy suggests the few-layered molecules preserve the temperature-induced spin-crossover switching observed in the bulk form but show a drastic increase in thermal hysteresis unique to these thin 2D molecule assemblies. The trapping of spin bistability with decreasing layer number can arise from domain wall dynamics in reduced dimensions. Our results establish molecular solids with strong anisotropy of intermolecular interactions as precursors to a novel class of 2D materials, affording new possibilities to control functionalities through substrate and interlayer interactions.

## Introduction

The family of van der Waals 2D materials has provided extraordinary opportunities to explore fundamental physics in reduced dimensions and advance research in nanomaterials and nanoscale devices[1,2]. A wide range of inorganic atomic 2D crystals[3-5] with drastically different physical properties have been investigated, revealing intriguing emergent phenomena at the few-layer and monolayer limits. Interlayer interactions allow additional functionality control to be achieved by stacking and twisting of homo- and hetero-layers of 2D crystals[6,7 1,8].

The search for molecular analogs of these 2D inorganic crystals, which provide precise chemical tunability, has attracted much attention in recent years[9-17]. Discrete molecules provide a diverse range of functionalities that can be tuned through synthetic modifications. In particular, spin-active molecules have been explored extensively in recent years vis-à-vis their potential applications in spintronic devices[18,19] and quantum information science[20-22]. Such molecular materials are conventionally considered as 0D systems with weak intermolecular interactions. Self-assembled molecular monolayers[23] are well established, but they assemble on specific substrates and typically exhibit molecular packing that differs from that of the bulk molecular crystals. Regular 2D single-crystalline packing of discrete molecules is particularly desirable because it provides homogeneity in molecular orientation and physical properties, but the experimental realization of such packing in the case of non-planar molecules has been challenging.

Here, we report the fabrication and characterization of 2D crystals of [Fe($t$Bu$_2$qsal)$_2$] (FTBQS) spin-crossover (SCO) molecular assemblies[24] down to the monolayer thickness. The molecular design of FTBQS is asymmetric, leading to a layered 2D molecular packing with strong π-π and σ-π van der Waals bonding for the in-plane direction and weak van der Waals interactions along the out-of-plane direction[24] (see Fig. 1a & b). Mechanical exfoliation is used to thin down the molecule layer number while preserving the same single-crystalline structure as that in the bulk crystals. With decreasing layer number, we observed an evident increase in the width of SCO transition thermal hysteresis from white light reflection contrast spectroscopy, implying the important roles of molecule-substrate interactions and dimensionality reduction.

## Results

The [Fe($t$Bu$_2$qsal)$_2$] (FTBQS) complex[24], where $t$Bu$_2$qsal$^-$ is a deprotonated form of 2,4-diterbutyl-6-((quinoline-8-ylimino)methyl)phenol ($t$Bu$_2$qsalH), crystallizes as a solvent-free molecular material in the monoclinic space group $P2_1/c$ (Fig. 1a). Molecular layers are formed parallel to the (100) plane (the x-y plane in Fig. 1b) due to efficient π-π and σ-π interactions between the quinoline fragments of the FTBQS molecules. while in the out-of-plane direction (the z-axis in Fig. 1b) these layers are coupled by weak van der Waal interactions between the $t$Bu group. The Fe(II) ions of this complex have been shown to exhibit spin bistability, undergoing SCO transition between the high-spin (HS, paramagnetic S=2) and low-spin (LS, diamagnetic S=0) states, as schematically shown in Fig. 4a. With decreasing temperature, the FTBQS bulk crystal undergoes the first-order HS → LS transition at 117 K and the reverse LS → HS transition at 129 K during the warmup process[24]. These transitions are accompanied by abrupt changes in the unit cell size due to the large difference in the average metal-to-ligand bond lengths between the LS state (1.949(2) Å for Fe–N bonds and 1.945(1) Å for Fe–O bonds) and the HS state (2.167(2) Å and 1.997(1) Å, respectively).

Mechanical exfoliation of the FTBQS bulk crystals is performed with the well-established Scotch tape method conventionally used in 2D atomic crystals fabrication. The (100) plane of the bulk single crystal is attached to the tape and deposited onto either Si/SiO$_2$ substrates or fused silica substrates. The exfoliated flakes are identified under an optical microscope through their white light color contrast and subsequently examined by atomic force microscopy (AFM) to determine the flake thickness and molecular layer numbers. The single-layer thickness is measured to be ~1.7 nm, in good agreement with the value of 1.8 nm estimated from the lattice constants (see Fig. S1 for multi-step layer height profile and Fig. 1c). We obtained exfoliated molecular layers down to the monolayer thickness, and the exfoliated flakes remained stable under ambient conditions for a few days. As shown in Fig. 1d, the lateral size of the exfoliated flakes ranges from a few microns to tens of microns, consistent with the sizes of the conventionally obtained 2D inorganic crystals. Thus, while the π-π and σ-π intermolecular interactions within the layer are non-covalent in nature, it is the higher strength of these interactions, as compared to the very weak van der Waals forces between the layers, that allows the exfoliation of the ultrathin crystalline flakes of these molecular solids. These exfoliated molecular flakes can be readily integrated with other 2D systems. Fig. 1e shows an example heterostructure of few-layer FTBQS and a monolayer MoS$_2$ built by dry transfer technique (see Method), which demonstrates the potential of building deterministic organic-inorganic hybrid heterostructures with van der Waals molecules.

To verify the material's stability at the few-layer limit, we measured the Raman spectroscopy as a function of the layer numbers. Figure 2a shows the Raman spectra of samples of different thicknesses under the same experimental conditions (see Methods). The spectra are shifted vertically for clarity and rescaled in intensity for comparison. The plotted high-frequency range (> 1000 cm$^{-1}$) in Fig. 2a has the strongest Raman signals and corresponds to vibrations of the ligands, and the full frequency range Raman spectrum shows a good agreement with the calculated Raman response for this complex, as discussed in the Supplementary Information (SI). Compared to the thick flakes (~30L) that can be approximated as bulk crystals, the Raman spectrum from the monolayer sample shows peak linewidths' broadening, relative intensity shift, and slight peak position shifts of up to 4 cm$^{-1}$. These Raman signal changes in the thin layer samples may be attributed to the sample degradation due to laser illumination or interactions between molecules and the substrate, which will require further theoretical investigation. The overall consistency in Raman features in samples of different thicknesses indicates that the molecular structure of these mechanically exfoliated flakes remains stable down to the monolayer limit.

The layer dependence of the visible light absorption in these exfoliated samples was measured by white light reflection contrast spectroscopy (see Methods). In the thin-film limit, the absorption is proportional to the white light reflection contrast spectrum through the Kramers-Kronig relations[25]. Fig. 2b shows the room-temperature reflection contrast spectra for samples on transparent fused silica substrates. The main absorption feature, located at ~ 2.3 eV, is assigned to a combination of the d-d transition at the Fe(II) site and the metal-to-ligand charge transfer (MLCT) transitions (see Fig. S5 and S6 in SI). We also observed weak PL, consistent with the optical absorption feature, as shown in Fig. S3. With decreasing layer number, the absorption amplitude decreased but with no obvious spectral change. This observation implies no significant layer dependence in the material's electronic structure, consistent with our expectations for these weakly interacting molecular complexes.

To further probe the spin bistability of the FTBQS flakes of different thicknesses, we have investigated the temperature dependence of their optical absorption. The HS ↔ LS transition leads to abrupt changes in the lattice constants, giving rise to substantial differences in the energies and amplitudes of the characteristic d-d and MLCT optical absorption bands (see SI), and such optical approach has been used to monitor spin transition in other SCO molecules[26-28] [29]. We first show the variation of bulk crystals' visible range absorption spectra during the warmup process at the rate of 3 K/min (Fig. 3a). The observed spectral differences between the HS state (red) and LS state (blue) can be qualitatively reproduced from calculations of the material's absorption coefficients (see SI). Figure 3e is the corresponding absorption amplitude measured at 535 nm (~2.32 eV) in a thermal cycle, and we can infer the HS↔LS transition from the absorption amplitude change. The measured transition temperature and thermal hysteresis are consistent with previous XRD and magnetic susceptibility measurements[24] in bulk crystals, confirming our assignment through optical spectroscopy. We note that the transition temperatures extracted here correspond to the cryostat thermal bath and can differ from the actual sample temperatures by a few Kelvins.

A strong dependence of the spin transition temperatures on the number of molecular layers has been observed in the exfoliated samples. Fig. 3 b-d show the reflection contrast (proportional to absorption in this experimental condition) spectra for 13L, 10L, and 5L samples taken at 0.3-0.5 K/min, and Fig. 3 f-h are the corresponding absorption amplitudes at 535 nm. The measured temperature range is 85-300 K to avoid possible complications from light-induced SCO at lower temperatures[26,30]. At ~85 K, i.e., below the bulk spin transition temperature, the spectra from these few-layer samples showed an obvious increase in the absorption amplitude at 535 nm (~2.32 eV) and an additional optical feature near 2.8 eV, which are similar to the observed LS spectrum in the bulk sample and indicative of the transition to the LS state. However, the absorption spectra changes in the exfoliated samples were more subtle than those in the bulk, especially at regions around 1.8 eV. In addition, the HS→LS conversion during the cooldown was more gradual when compared to the bulk. These differences may result from the possible coexistence of HS and LS states in exfoliated samples, which has been reported in amorphous or polycrystalline SCO thin films grown by solution or vapor-phase deposition[31]. In contrast, the reverse LS→HS transition was much better defined by the abrupt absorption amplitude decrease during the warmup (see Fig.3f-h grayed areas). The extracted LS→HS transition temperatures T$_{LS→HS}$ are plotted as a function of layer numbers in Fig.3j, showing a ~170 K increase from the bulk transition temperature when the bulk crystal is thinned down to a few layers. In particular, the transition in the 5L sample is only achieved after an additional hour of wait time at room temperature. Data obtained from multiple thermal cycles are consistent and the thickness dependence of the warm up LS→HS transition is summarized in Fig. 4b. We also varied the temperature change rate down to 0.1 K/min in several measurements and did not observe any obvious differences compared to the data presented in Fig. 3 and 4. Thus, our results reproducibly demonstrate a drastic increase in the $T_{LS \to HS}$ and the remarkable broadening of thermal hysteresis in thin exfoliated samples.

## Discussion

To understand such layer dependence of the spin-switching properties, we consider the possible contributions from the external factors such as substrate interactions and the intrinsic mechanism due to the reduction in dimensionality. The HS↔LS transition hear is known to lead to substantial unit cell parameter changes[24]. Because the exfoliated few-layer samples are conformed to the substrates, the system needs to overcome additional substrate strain[32] during the spin state switching. Thinner samples have a higher surface-to-volume ratio, leading to stronger strain confinement from the substrate. We, therefore, expect an increase in the HS and LS transition energy barrier for samples with decreasing layer number, consistent with our observation. On the other hand, the dimensionality reduction can also modify this first-order phase transition[33]. When considering the phase transition kinetics, the nucleation and growth of a new phase will be distinctly different for samples with stronger 2D confinement, where the phase domain walls will be 1D. The reduction in the available domain propagation phase space in 2D samples may also lead to significantly longer transition times[34] and contribute to the observed large thermal hysteresis. While most SCO thin films have been reported to have reduced spin-switching cooperativity and smaller thermal hysteresis[31], we note that the single-crystalline nature of the investigated ultrathin samples ensures much lower densities of crystal imperfections compared to the conventionally studied amorphous and polycrystalline SCO films. As the SCO transitions are expected to preferentially nucleate at defect and surface sites [35,36], the exfoliated layers can still preserve cooperativity at the few-layer limit. A comprehensive picture will still require further theoretical investigation.

In conclusion, we obtained the single-crystalline, 2D FTBQS molecular layers at few-layer and monolayer thickness using the mechanical exfoliation method. The molecules remain intact, showing consistent Raman signals and absorption down to the monolayer. The SCO transition, on the other hand, shows a drastic thickness dependence as revealed by reflection-contrast spectroscopy: as the layer number decrease, the thermal hysteresis range increases more than 100K. It demonstrates the efficient tuning of the molecules' magnetic properties through layer number and substrate interactions and is of fundamental importance to understand the SCO processes in 2D. Our findings establish a new paradigm for the preparation of molecular 2D materials, which can be readily integrated into 2D spintronic devices[15,17] to incorporate such molecular properties as spin bistability. Through different molecular designs, additional functionalities are also possible with this new type of van der Waals molecule-based materials to allow 2D integration of other magnetic molecules, such as arrays of single-molecule magnets or molecular spin qubits.

## Methods

**Growth of FTBQS bulk crystal.**
The procedure to synthesize the single crystal of [Fe(tBu2qsal)$_2$] used for mechanical exfoliation is similar to that described in the previous work[24].
*(a) Starting Materials:* Iron(II) perchlorate hexahydrate (98%), triethylamine (Et$_3$N) (99%), and formic acid (98%) were obtained from Millipore Sigma, 8-aminoquinoline (>97%) from Matrix Scientific, and 3,5-ditertbutylsalicylaldehyde (>98%) from TCI. All reagents were used as received. Methanol and acetonitrile (MeCN) were obtained as anhydrous HPLC grade solvents from Millipore Sigma. Anhydrous MeCN was further purified by passing through a system of double-drying columns packed with activated alumina and molecular sieves (Pure Process Technologies). Anhydrous acetone ($Me_2CO$) was obtained from Acros Organics. Elemental analysis was performed by Atlantic Microlab, Inc. (Norcross, GA, USA).

*(b) 2,4-di(tert-butyl)-6-((quinoline-8-ylimino)methyl)phenol (tBu2qsalH):* 8-aminoquinoline (0.72 g, 5 mmol) and 3,5-ditertbutylsalicylaldehyde (1.17 g, 5 mmol) were placed in a round bottom flask. 29.5 mL of MeOH and 3 drops of formic acid were added, and the solution was stirred for 1 h under reflux. The orange solid that precipitated was isolated by filtration, washed with small amount of MeOH, and dried by suction. Yield = 0.82 g (45.6%).

*(c) [Fe(tBu2qsal)$_2$]:* The reaction was performed under nitrogen using standard Schlenk techniques. A solution of $Fe(ClO_4)_2 \cdot 6H_2O$ (272 mg, 0.75 mmol) in 15 mL of MeCN was added to a solution of tBu2qsalH (540 mg, 1.5 mmol) and $Et_3N$ (209 μL, 1.5 mmol) in 10 mL of $Me_2CO$, and the mixture was left undisturbed. Big plate-shaped dark-green crystals that formed after 1 day were recovered by filtration, washed with a small amount of acetone, and dried by suction. Yield = 452 mg (77.8 %). *Elem. analysis:* Calcd. (Found) for $FeO_2N_4C_{48}H_{56}O$, %: C 74.40 (74.37), H 7.23 (6.98), N 7.02 (7.31).

**Mechanical exfoliation of 2D FTBQS and fabrication of heterostructure.**

As described in the main text, flakes of FTBQS were mechanically exfoliated from the bulk crystals onto $SiO_2$/Si or fused silica substrates, using the Scotch tape method. The studied heterostructure is composed of FTBQS, and $MoS_2$. The $MoS_2$ (hq graphene) were exfoliated from their bulk crystal onto transparent elastomer stamps (poly dimethyl siloxane, PDMS). The heterostructure was then assembled by the layer-by-layer dry transfer technique[37].

**Optical spectroscopies**

*(a)* Raman spectroscopy: A 532 nm laser source is used as excitation, filtered by BragGrate™ Spatial Filter. The laser was focused on the sample with a focal spot diameter ~ 1 μm using an objective. The incident power is ~10 μW. The reflected light was collected, filtered by a long-pass filter (Semrock RazorEdge), and focused into the spectrometer and CCD (Princeton Instruments SpectraPro HRS300 + PIXIS) for detection.

*(b)* White light reflection contrast spectroscopy (for exfoliated samples): A high-intensity halogen lamp was used as the white light source. The focal spot size is ~ 1 μm. The reflection contrast spectrum $\Delta R/R$ was obtained by comparing the reflectance from the FTBQS sample ($R'$) and that of the bare substrate ($R$): $\Delta R/R = (R' - R)/R$.

*(c)* Optical absorption spectroscopy (for bulk crystals): Optical absorption spectra were recorded on a double-beam Cary 300 Bio spectrometer (Agilent). The samples were prepared from small single crystals that were thinned down with a Scotch tape 3-4 times and mounted on a copper plate so as to completely cover a small aperture of ~1 mm diameter. The plate was inserted into a closed-cycle He cryostat (Janis) capable of reaching 10 K. The spectra were recorded in the transmission mode in the range from 380 to 800 nm (3.26 to 1.55 eV).

**Author Information**

**Corresponding Author**

*Correspondence to: xxzhang@ufl.edu  shatruk@chem.fsu.edu

**Author contributions**

X.Z. and M. S. designed the study. J.K. fabricated the 2D samples and heterostructures, and performed the measurement on 2D samples, assisted by X.Z. M.G. grew the FTBQS bulk crystals and performed characterization on bulk samples. J.X. assisted the AFM measurements of 2D samples. D.L. performed the theoretical analysis of optical absorption coefficient under the guidance of T.S.R.. J.J., R.Z. performed the theoretical calculation of Raman spectroscopy under the guidance of H.-P. Z.. All authors contributed to writing the manuscript and interpretation of the results.

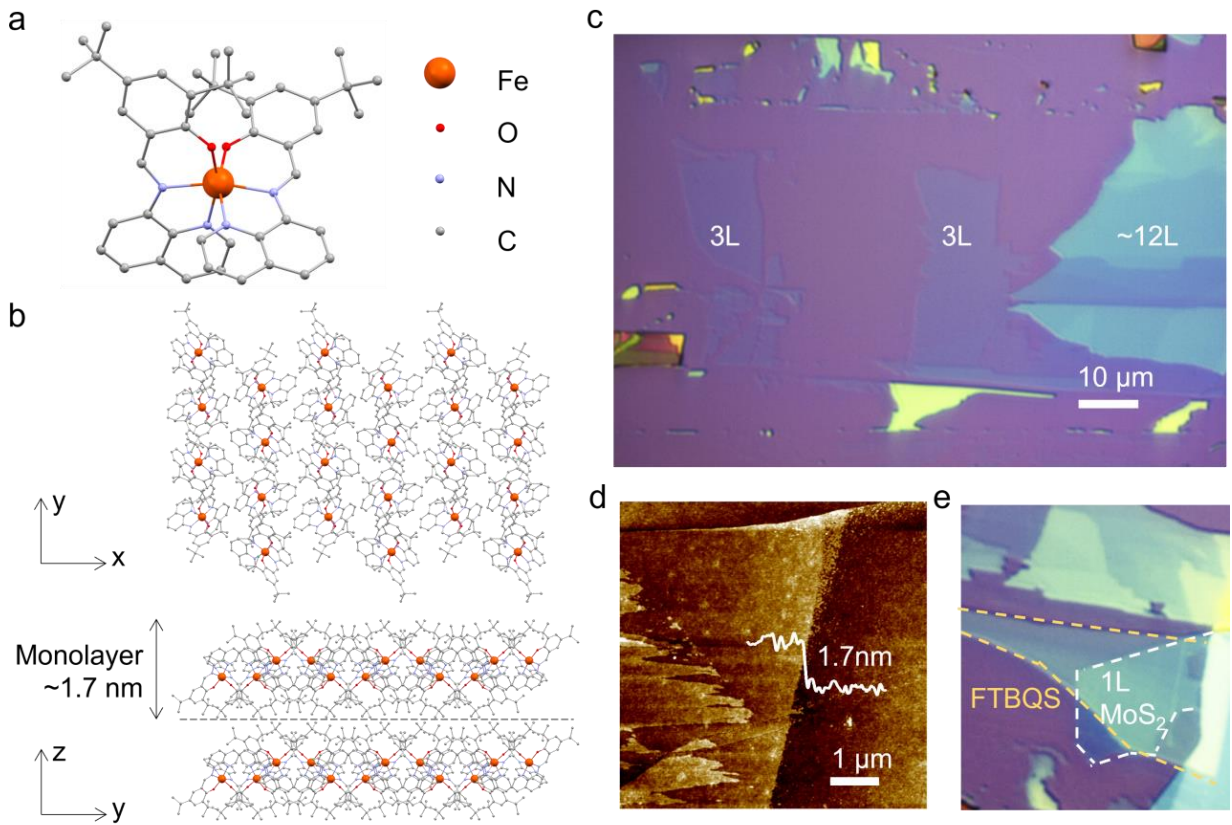

**Figure. 1 Molecular structure and identification of exfoliated [Fe(*t*Bu$_2$qsal)$_2$] layers. a,** The crystal structure of [Fe(*t*Bu$_2$qsal)$_2$] (FTBQS) molecule complex. The H atoms are omitted for clarity. **b,** The single-crystalline structure of FTBQS. The x-y plane corresponds to the (100) plane of the crystal. The *z*-direction corresponds to the out-of-plane direction, and the unit cell thickness is measured to be ~1.7 nm. The color code follows the notations in **a**. **c,** The white light microscope picture of single-crystalline FTBQS flakes of different layer numbers. The layer number is assigned from the AFM data. **d,** The atomic force microscope (AFM) image of the monolayer FTBQS region on SiO$_2$/Si substrate. The height profile, which shows a 1.7nm monolayer step, is shown. **e,** The microscope image of the assembled heterostructure assembled through dry transfer technique. The yellow and white dashed lines outline the crystal edges of ~10L FTBQS and a monolayer MoS$_2$.

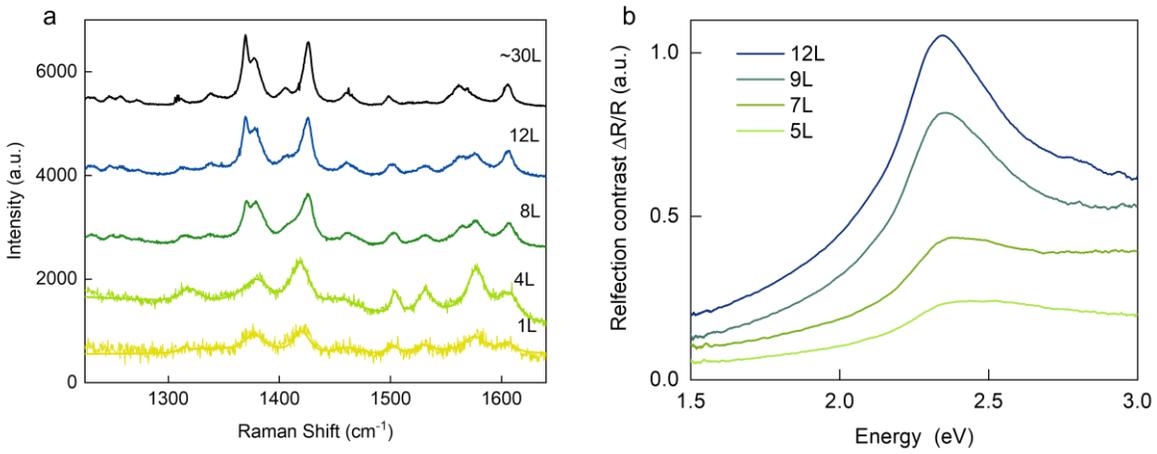

**Figure. 2 Raman and reflection contrast spectroscopies of exfoliated FTBQS samples. a,** Raman spectra as a function of the layer number of samples. 532nm laser was used as the excitation source. The plotted region has the strongest Raman signal. The Raman spectrum down to ~120 cm$^{-1}$ is included in the SI. **b,** White light reflection contrast spectroscopies as a function of the layer number. Samples for this measurement were exfoliated onto transparent, fused silica substrates.

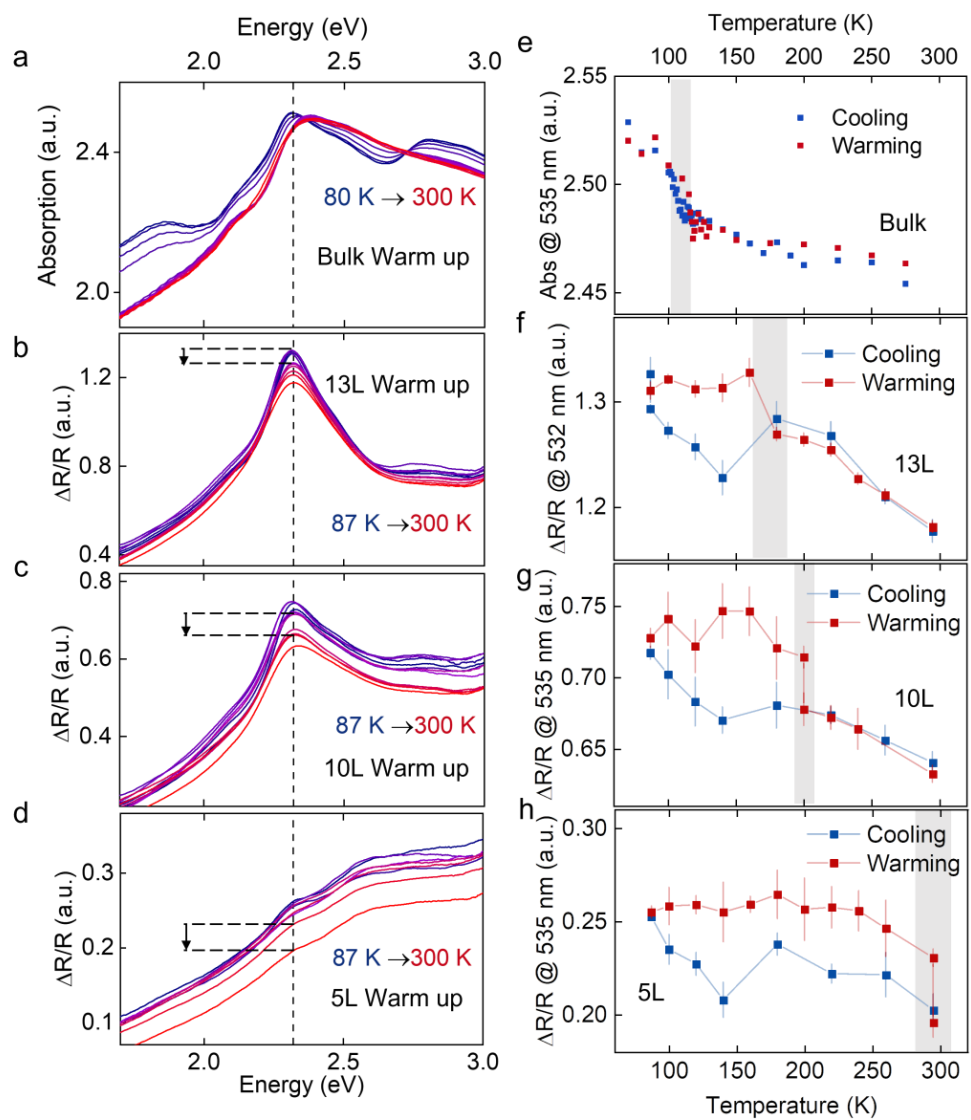

**Figure. 3 Spin-crossover transitions measured by absorption and reflection contrast spectroscopies.
a-d,** Absorption for the bulk sample and reflection contrast spectra for 13L, 10L, and 5L samples, respectively, during the warmup process. The arrow in indicates the abrupt amplitude change that is attributed to the LS→HS transition. The abrupt change in **c** occurs after 0.2 hr wait time at 220k, and that in **d** occurs after 1 hr wait time at room temperature. **e-h,** The absorption/reflection contrast amplitudes were taken at 535nm (as indicated by the dashed vertical lines in **a-d**) in a cooldown-warmup thermal cycle. The grey area indicates the extracted LS→HS transition temperature range.

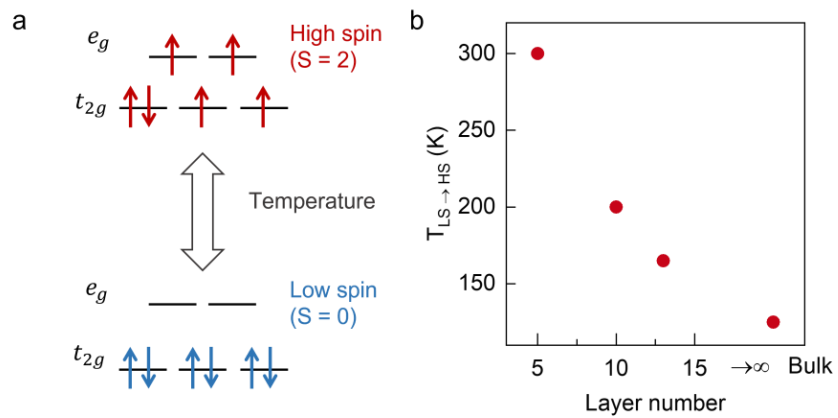

**Figure. 4 Thickness-dependent spin-crossover transition temperature in exfoliated FTBQS samples.**
**a.** Schematic illustration of the spin configuration of the Fe(II) ion for the high spin (HS) and low spin (LS) states. **b**, The layer dependence of the measured LS→HS transition temperature from Fig. 3. The transition temperature is extracted as the abrupt changes in reflection contrast, as the greyed out area in Fig. 3 e-h.